\documentclass[amsmat,amssymb,amsfonts,aps,prb,twocolumn,showpacs]{revtex4}

\usepackage{graphicx}
\usepackage{dcolumn}
\usepackage{bm}
\begin{document}

\title{Single exciton spectroscopy of semimagnetic quantum dots }

\author{J. Fern\'andez-Rossier}
\affiliation{Instituto Universitario de Materiales de Alicante,
 Universidad de Alicante, San Vicente del Raspeig, 03690 Spain }

\date{\today}

\begin{abstract}
A photo-excited  II-VI semiconductor nanocrystal  doped with a
few Mn spins is considered. The effects of spin-exciton interactions and the
resulting multi-spin correlations on  the photoluminescence are calculated by
numerical diagonalization of the   Hamiltonian,  including exchange interaction
between electrons, holes and Mn spins, as well as spin-orbit interaction.
 The results provide a unified description of recent experiments
 of photoluminesnce of dots with one and many Mn atoms as well as 
  optically induced
ferromagnetism in semimagnetic nanocrystals.

\end{abstract}

 \maketitle

Control of the wave function of spins embedded in semiconducting materials is
in the road-map of the quantum hardware development \cite{Kane}.  Optical
excitation and probing of single exciton confined in a nanometric region
permits to manipulate the quantum state of  the exciton spin 
\cite{Bonadeo,Li}, or   the spin degrees of freedom coupled to the excitons
like  nuclei  \cite{Nuclei},  donor electrons \cite{Nuclei,Merlin,Lu}  or Mn
ions  \cite{Merlin,Besombes,Besombes2, Besombes3,
Besombes4,Maksimov,Bacher,Doro,Smith}. Recent reports
\cite{Besombes,Besombes2,Besombes3} of optical detection of the
photoluminescence (PL) of a single nanocrystal of II-VI semiconductor doped
with a {\em single} Mn atom showed evidence of a one on one correspondence
between the energy of the emitted photons  and the quantum state of the Mn spin
after photon emission.  Substitutional Mn in (II,Mn)VI semiconductors is a
neutral impurity\cite{Furdyna} with 5 electrons in the open $d$ shell that
behave  like a spin $S=\frac{5}{2}$. Therefore, the experiment of ref.
(\onlinecite{Besombes}) is a proof of principle  of the optical manipulation
and detection of the quantum state of a single spin $S=5/2$.

Several other groups have reported the fabrication and optical spectroscopy of 
single  nanocrystals of II-VI semiconductors doped with  tens of Mn
atoms\cite{Maksimov,Bacher,Doro,Smith,Nature2005}.  The PL   spectrum of one of these
semimagnetic nanocrystal  is broader than that of a pure
dot\cite{Besombes,Bacher},  and it shows strong sensitivity to the
application of  magnetic fields \cite{Bacher,Doro}, as a result of the exchange
interaction of  conduction band (CB)  electrons and valence band (VB)
holes with  the Mn spins. 
Here,  a  quantum theory of the single exciton spectroscopy of a
nanocrystal with a few Mn spins is presented. The goal is to provide a 
unified description of the PL spectra of nanocrystals
 with one \cite{Besombes,Besombes2,Besombes3} and many Mn atoms,
 with emphasis on how to extract information about the quantum state of
 the Mn spins from the PL spectra.
 With that aim, the standard Hamiltonian for semimagnetic nanocrystals  is
{\em solved exactly}  for $N_{Mn}=$1,2,3 and 4   Mn ions both
when one exciton is present (excited state manifold, XMS) and
absent ( ground state  manifold, GSM). The  optical transition rates between
the GSM and XSM  are calculated and  PL  spectra are obtained both in the
Faraday and Voigt configurations for a range of situations. Our results
agree very well with the PL experiments with dots with one 
Mn  \cite{Besombes,Besombes2,Besombes3} and many  \cite{Bacher,Doro}.
Previous theory work reporting exact diagonalizations
 \cite{Govorov,Bhatta} have addressed single Mn dots with 
  spherical shape, resulting in PL spectra different from those obtained here
  and reported experimentally \cite{Besombes,Besombes2}.

{\em Hamiltonian}.
The standard Hamiltonian \cite{Furdyna,Efros,Govorov,Bhatta,JFR04} for diluted magnetic
semiconductors  describes   CB electrons and VB  holes  interacting with
localized Mn spins ($\vec{M}_I$) via a local exchange interaction  and coupling to an external
magnetic field, $\vec{B}$. For the results presented here  the  short range
antiferromagnetic superexchange is not effective\cite{JFR04}. Both  CB electrons
and VB holes are confined in all the 3 spacial  directions resulting in a
discrete single particle spectrum.   The  Hamiltonian  reads ${\cal H}={\cal
H}_0+{\cal H}_1+{\cal H}_{eh}$,  where ${\cal
H}_0=-g\mu_B\vec{B}\sum_I\vec{M}_I$ and 
\begin{eqnarray}
{\cal H}_1= \sum_{n,\sigma,\sigma'}\left(\epsilon^{e}_{n}\delta_{\sigma,\sigma'}
-g_e \mu_B \vec{B} \cdot\frac{\vec{\tau}_{\sigma,\sigma'}}{2} \right)
  c^{\dagger}_{n,\sigma}
  c_{n,\sigma'} +
\nonumber \\ 
\sum_{\nu}\epsilon^{h}_{\nu}(\vec{B})   d^{\dagger}_{\nu}
  d_{\nu} 
+
 \sum_{I}
\vec{M}_I\cdot \left(J_e { \vec{S}}_e(\vec{r}_I) 
+J_h  { \vec S}_h(\vec{r}_I) \right)
\label{hamil}
\end{eqnarray}
 In the absence of electron-hole ({\em eh}) pairs, 
the $\vec{M}_I$ interact only with $\vec{B}$.  
The first term in ${\cal H}_1$  describes the CB electrons confined in the
nanocrystal orbital levels $\psi_{n}(\vec{r})$ and coupled to the external magnetic
field. The Pauli matrices are denoted by  $\vec{\tau}_{\sigma,\sigma'}$
 and  
$  c^{\dagger}_{n,\sigma}$  stands for the CB electron creation operator. 
The second term describes the VB holes confined in the nanocrystal, including
the spin orbit interaction \cite{dot-holes,Kohn-Luttinger};
 the operator $  d^{\dagger}_{\nu}$ creates
a  VB hole in the spin-orbital   $\phi_{\nu}(\vec{r})$.
The third and fourth term describe the exchange between Mn and CB electrons
and VB holes, respectively. 
Both the CB-Mn and the VB-Mn exchange 
are local and couple the carrier spin density at the Mn location
$  \vec{S}_{e,h}(\vec{r}_I)$ to 
 $\vec{M}_I$.
 The {\em eh} exchange is described by ${\cal
H}_{eh}=-J_{eh}\vec{S}_e\cdot\vec{S}_h$. 
Direct {\em eh} Coulomb interaction and orbital magnetism  are irrelevant in
the small dots considered here\cite{JFR04}.

We consider  dots whose single particle energy spacing is much larger than all
the other intra-band energy scales of the Hamiltonian. The shape of the dot 
determines the degree of light hole (LH) heavy hole (HH) mixing of the VB
single particle states which, in turn, determines the relative importance of
the Ising and spin flip part of the hole-Mn and hole-electron exchange
interactions.   A  hard wall cubic  potential with dimensions
$L_x,L_y,L_z$ conveniently  describes the
interplay between the shape of the dot and the single-hole states \cite{dot-holes},
 although real dots probably have  more complicated geometries. 
We always take $L_z<L_x,L_y$ and we consider light propagation along the  $z$
axis. The CB  electron envelope functions are  given by \cite{JFR04}
 $\psi_{\vec{n}}\equiv \sqrt{\frac{8}{V}} Sin(k_x x) Sin(k_y
y) Sin(k_z z)$ where $k_i\equiv \frac{ \pi n_i}{L_i}$, $i=x,y,z$, $V=L_x L_y
L_z$ and $n_i$ are
positive integers. The description of the confined VB single particle states 
follows  that of reference (\onlinecite{dot-holes}): a zero dimensional
version  of the  6 band Kohn-Luttinger Hamiltonian 
\cite{Kohn-Luttinger}. The envelope states for the single particle
lowest energy doublet are
$\psi_0(\vec{r})\equiv\psi_{1,1,1}(\vec{r})$, both for VB holes
and CB electrons.  The wave functions of the VB lowest energy doublet
 are   $|+\rangle= c_1 |\frac{3}{2},\frac{+3}{2}\rangle +c_2
|\frac{3}{2},\frac{-1}{2}\rangle$ and  $|-\rangle={\cal K}|+\rangle $ where
${\cal K}$ is the time reversal operation and  $|J,M\rangle$ are the standard
\cite{Kohn-Luttinger} light hole (LH)  and heavy hole (HH)
states at the $\Gamma_8$ point.

 If $L_x=L_y$ LH-HH mixing vanishes ($c_2=0$) and so they do the
 matrix elements $\langle \pm |S_h^{x,y}| \mp\rangle$,  
 resulting in a pure Ising ($zz$)  hole-Mn and $eh$ exchange coupling.
 For $0<|L_x-L_y|<L_z$  
there is some HH-LH mixing ($0<|c_2|<|c_1|$) that, on top of
the dominant Ising coupling,  permits  some spin flip ($xx$, $yy$) 
hole-Mn and $eh$  interaction. 
In contrast with the VB holes, the
CB electron-Mn interaction is perfectly isotropic \cite{JFR04}. 
The results shown here belong to 4 different dots with $(L_x,L_y,L_z)$
(7,7,4) nm (dot $\#1$), (15,15,3) nm (dot $\#2$), (7,5,2) nm (dot $\#3$) and
(7,3.7,2) (dot $\#4$).
The strength of the exchange coupling between the carriers and the
Mn spins is 
$j_{e,h}(I)\equiv J_{e,h} |\psi_0(\vec{r}_I)|^2$ .
For a given dot the  mean  value of the exchange coupling is
$\langle j_{e,h}\rangle=\frac{J_{e,h}}{V}$ and the maximum
value is $8 \langle j_{e,h}\rangle$. 
For CdTe we have\cite{Furdyna} $J_h$=+60 meV nm$^3$ and
$J_e=-15$meV nm$^3$.  We take $g_e=-1.5$ and $g_h=-0.1$\cite{Besombes}. 
In order to establish the influence of the different exchange couplings
in the PL spectra of dots with 1 Mn, we first take
$J_{eh}=0$ (fig. 1) and then we study the effect of $eh$ exchange
(fig.  2a,b,c)

The many body states are classified according to the number of excitons
($N_X=0$ for the GSM and $N_X=1$ for the XSM) and to the number of Mn ions,
$N_{Mn}$. The number of states in the GSM  of a given nanocrystal is 
$(6)^{N_{Mn}}$.  If only the lowest energy doublet of both the CB and VB states
are kept, there are four possible exciton states, labeled by $|\pm 1\rangle_X$
and  $|\pm 2\rangle_X$ following their total angular momentum, and  the number
of states in the XSM is $4\times (6)^{N_{Mn}}$.  The many-body states are found
by standard numerical diagonalization of the Schrodinger equations ${\cal H}_0
|G\rangle=E_G |G\rangle$ and $\left({\cal H}_0 +{\cal H}_1\right)|X\rangle=E_X
|X\rangle$. The GSM states only feature Mn spin  coordinates whereas the XSM
states include  Mn spins,  as well as 1 CB electron and 1 VB hole that occupy
linear combinations of the single particle states described above.

In the absence of magnetic field and superexchange all the GSM states are 
degenerate. In contrast, Ising hole-Mn coupling of dot $\#1$ lifts splits the
24 levels in 6 quartets, corresponding to the six possible relative
orientations between the Mn and the hole spin along the $z$ direction. Both
electron-Mn and electron-hole exchange coupling further split the 6 quartets
into  12 doublets.  Depending on the relative size of $J_{eh}$ and $J_{e}$ 
the lowest energy state of the XSM is either 
  $|\pm 1\rangle_X \otimes  |\mp\frac{5}{2}\rangle$ 
  or $|\pm 2\rangle_X \otimes  |\mp\frac{5}{2}\rangle$.
The diagonalizations  show that  
 the lowest energy states of the XSM of dots with $N_M=$2,3,4   describe 
 fully polarized Mn spins along $L_z$.
Hence, the exact many-spin wave functions of the Mn coupled to zero
dimensional exciton feature ferromagnetic correlations, very much like
bulk carrier mediated ferromagnetic order
  \cite{DMS-bulk}
and in agreement with
experimental observations in photoexcited nanocrystals\cite{Smith}.
   
Transitions from the XSM to the GSM are  possible via spontaneous emission
of a photon with the adequate energy $\hbar \omega$ and polarization. We 
consider circularly polarized photons.
The rate of spontaneous emission of a photon of energy 
$\hbar\omega=E_X-E_G$ 
  from the state $|X\rangle$ to the
state $|G\rangle$ is given by the Weisskopf-Wigner expression
$\Gamma_{GX}=
 \frac{4 \omega^3}{4\pi\epsilon_0\time 3 \hbar c^3} 
 |\langle G| {\cal P}^{\pm} |X\rangle|^2 $
where
${\cal P}^{\pm}=\sum_{\nu,n,\sigma}  \langle \nu|e(x\pm iy)|n\sigma\rangle
d_{\nu}^{\dagger}c_{n,\sigma}^{\dagger} +{\rm h. c.} $  
is the second quantization representation of the interband electric dipole
operator that yields the standard optical selection rules. 
Standard optical selection rules forbid photon
emission from $|\pm 2\rangle_X$ states.   Since the electric dipole operator
does not affect the Mn $d$ electrons,
the Mn spin part of the collective wave function
{\em does not change during the recombination process}. As a result,
most of the GSM to XSM
transitions that are allowed by the standard optical selection rule but
are forbidden due to the orthogonality of the spin part of GSM and the XSM
states.  This optical spin blockade, reminiscent of the spin blockade well
characterized in single electron transport \cite{Juanjo},  implies a huge
limitation to the otherwise large set  of XSM to GSM transitions.
The PL spectrum  is calculated by statistical weighting of the initial states
$|X\rangle$:
$${\cal I}_{PL}(\omega)=\sum_{X,G} P(X)  \Gamma_{GX}
\delta\left(\hbar\omega-(E_X-E_G)\right)$$
where $P(X)$ is the occupation probability of the XSM state $|X\rangle$. 
In general the calculation of $P(X)$ would involve the determination of the
non-equilibrium steady state density matrix \cite{Efros,Govorov}.  The experimental
results of  reference \onlinecite{Besombes} support approximating   $P(X)$ as
a  thermal distribution function with an effective temperature. 

\begin{figure}
[hbt]
\includegraphics[width=2.8in]{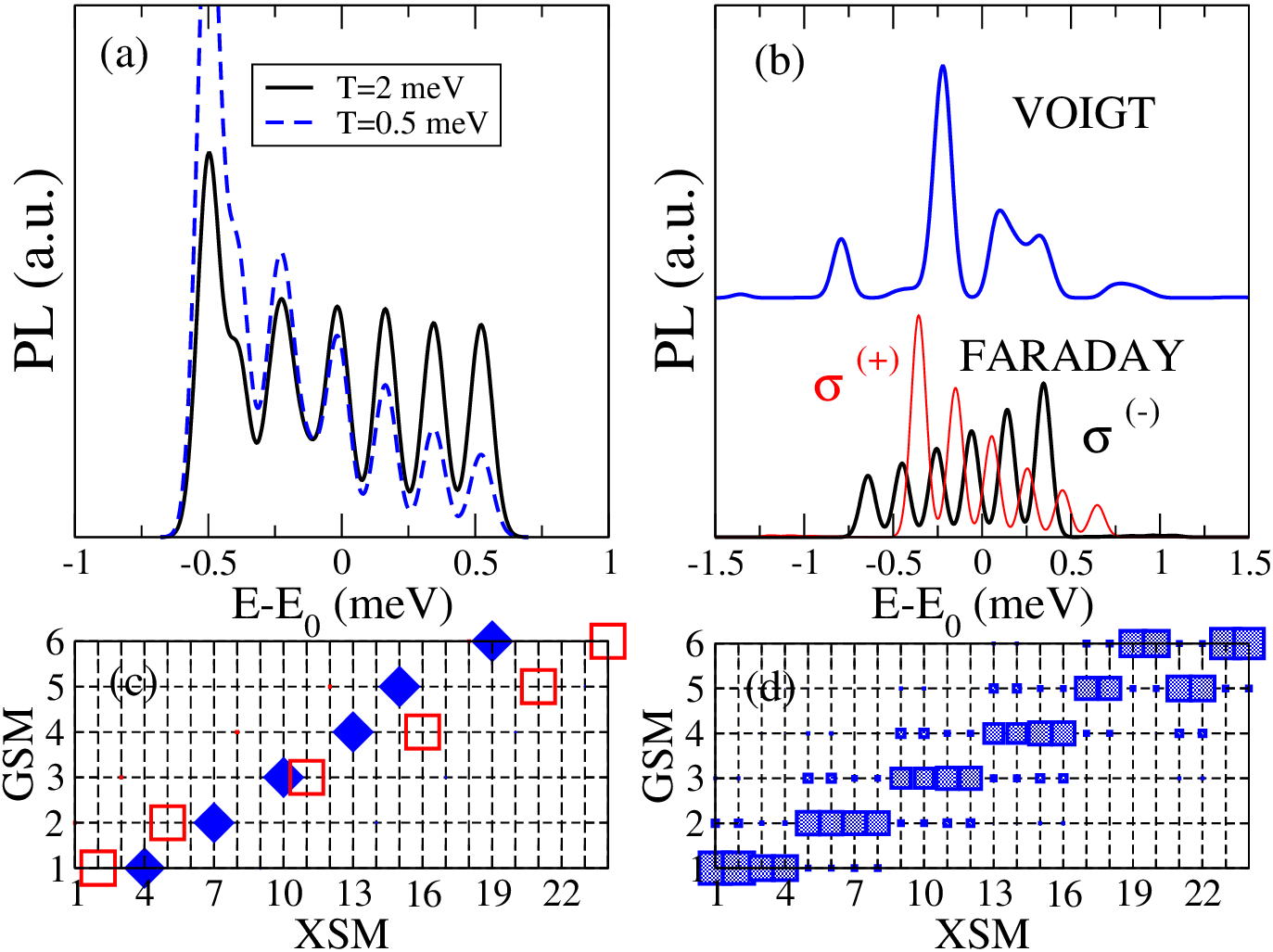}
\caption{ \label{fig1}(Color online).
(a)  PL for  dot $\#1$ with 1 Mn atom and two values
of T.   (b) Magneto PL
  in the Faraday ($(0,0,5T)$) and Voigt ($(5T,0,0)$)
configurations. (c,d) $\Gamma_{GX}$ for the Faraday (c) and Vo.igt (d) cases. } 
\end{figure}

In the following we show PL spectra  calculated following the
above recipe. The spectral lines are broadened using gaussians with a
FWHM of either $\gamma=$50$\mu$eV or $\gamma=$400$\mu$eV,  
adapted to the experimental
photodetector resolution of references 
(\onlinecite{Besombes}) and (\onlinecite{Bacher}) respectively. 
Single spin optical spectroscopy  calculations are shown in fig. 1. The zero
field  PL for  nanocrystal $\#1$ with a single Mn impurity and 2 effective
temperatures , 0.5 and 2 meV (6 and 24 Kelvin) is shown in fig. 1a.  The model 
accounts quantitatively for the experimental observations  \cite{Besombes}: 
only 6 lines are clearly seen, in spite of the fact that the XSM has
12  non degenerate lines, that occupy a spectral range of 1 meV. 
The  number 
of the lines differ from those of previous theory work
 (\cite{Govorov,Bhatta})
due to the different LH-HH mixing of their model.
The 6 lines undergo a polarization dependent splitting (fig. 1b, lower
spectrum) upon application of a magnetic field in the Faraday configuration,
$\vec{B}=(0,0,B_z)$, in  good agreement with the experiments
\cite{Besombes,Besombes4}. 

The  detection of a photon with a given polarization and energy in one of these
6 peaks yields information about the final state of the GSM in the optical
transition. That information is given, for the Faraday spectrum of fig. 1b, in
the fig. 1c. There the size of the symbols is proportional to 
$\Gamma_{GX}$  for $\sigma_-$ (open squares circles) and $\sigma_+$ (solid
diamonds).   It is seen how  each member of the $XSM$ is coupled via a given
circularly polarized photon with, at most, one member of the GSM. 
This fact {\em  permits to map the photon state (energy and polarization) to the
final Mn state} and is the second important result of this paper. In
particular,  going from low to high energy the six peaks from the  PL in the
Faraday case correspond to final Mn states going from $S_z=+\frac{5}{2}$ to 
$S_z=-\frac{5}{2}$. The zero field case has the same one-on-one correspondence.

The calculated PL in the Voigt configuration, $\vec{B}=(B_x,0,0)$, is
remarkably different. The  shape of the PL is modified drastically, and the
spectrum is much wider. The GSM states are the eigenstates of the $S_X$ spin
operator, whereas the XSM states have strong overlap with the eigenstates of
the $S_z$ operator, because the spin of the hole is pinned in that direction
and the hole-Mn  Ising coupling is dominant. The matrix $\langle G |{\cal
P}|X\rangle$ is now proportional to $\langle S_x|S_z\rangle$ making spin
blockade much less efficient, as shown in fig. 1d. This is seen in figure 1d
and accounts for the PL shape reported in \onlinecite{Besombes4}. 


We now consider the effect of the $eh$ exchange interaction, absent in the
results of figure 1. The Ising part of the $eh$ interaction splits bright and
dark excitons, whereas the spin-flip part of the $eh$ interaction, which is
only possible if there is LH-HH mixing, mixes $+1$ and $-1$ excitons resulting
into {\em linearly polarized} spectra \cite{Bonadeo}.
 In the presence of Mn, this
mixing competes with the Ising hole-Mn coupling.
The combined action of electron-Mn exchange and 
the transverse hole-Mn and $eh$ interactions
  mix somewhat bright and dark excitons. 
 In figure 2a-2c we show the PL
spectra corresponding to dot $\# 3$ (2a) and dot $\# 4$ (2b,2c),
 all of them with a single Mn atom and $J_{eh}=1$ meV. Dot $\# 3$
features some LH-HH mixing. Ising $eh$ interaction red shifts the  dark
excitons, resulting in the appearance of 3 low energy  peaks (compared with
fig. 1). The middle PL corresponds to a  Mn weakly coupled ($j_h=0.16 meV$,
$j_e=0.25 j_h$) to the exciton and a strong LH-HH mixing. The PL is linearly
polarized and the peak structure  of figs. 1a  is not resolved. The lower panel
corresponds to the same dot with a strongly coupled Mn ($j_h=0.31$ meV). 
There the 6 peak
structure is recovered, but the PL is linearly polarized, mostly in the central
peaks for which the hole-Mn Ising coupling is smaller.  These whole picture is
consistent and qualitatively identical to the experiments recently published
\cite{Besombes3}.

We finally address the central theme of this paper: how the PL spectra of the
dot evolve as the number of Mn atoms increases.  In this regard, we start
considering the a dot with 3 Mn atoms,    two of which are identically coupled
($j_h(1)=j_h(2)=0.38$ meV) and the third changes, so that  the relative
coupling to the fermions  defined by
$r=\frac{|\psi_0(\vec{r}_3)|^2}{|\psi_0(\vec{r}_{1,2})|}$ is varied 
from  from 1 to 0.1 (figs 2d to 2g).   The  symmetric case $r=1$  
shows a clean PL spectrum  with 16 lines corresponding to quantum states with
the $2S+1$ possible orientations  of the collective $S_Z$  of the 3 Mn atoms
that behave like a single magnet with ($S=15/2$).  These 16 peaks are still
seen for $r=0.85$. At   $r=0.5$ the PL is a rather featurless wide peak. When
$r=0.1$ the PL spectrum features 11 peaks, that are the analog of figure 2d for
2 Mn atoms behaving collectively with $S=10/2$. It is apparent that the weakly
coupled Mn spin does not affect the PL. 

\begin{figure}
[t]
\includegraphics[width=2.8in]{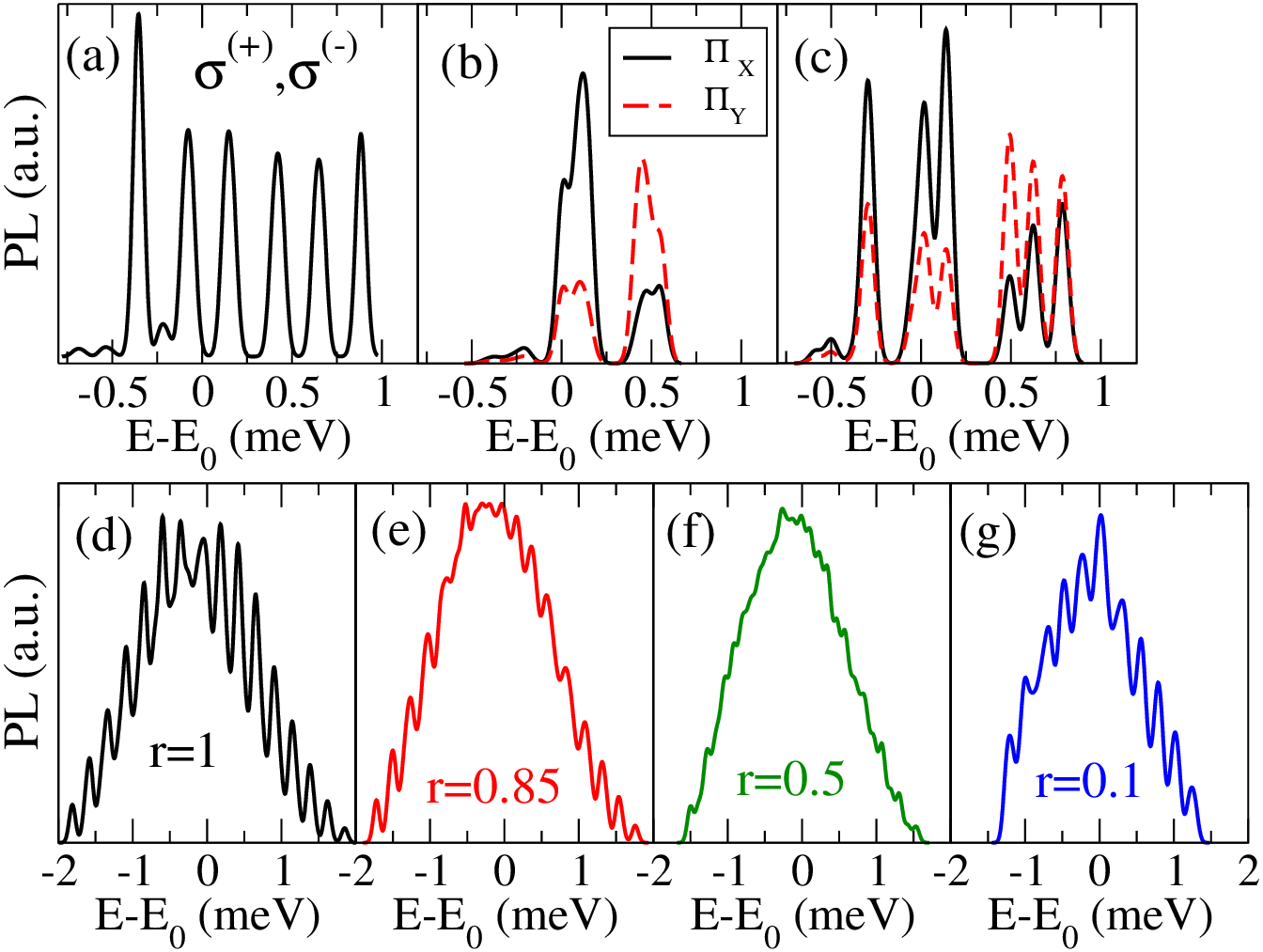}
\caption{ \label{fig2}(Color online).
 Upper panels: Influence of the
LL-HH mixing and the $eh$ exchange on the PL of single Mn PL for dots
  $\#3$ (2a) and $\#4$ (2b and 2c).  
2d-2g  Zero field PL spectrum for dot $\#2$ with 3 Mn impurities
as a function of the relative Mn-carrier couplings.}
\end{figure}

The results of figure 2d-2g suggest that a few Mn atoms, strongly coupled to the
exciton could be responsible of the main features of the PL signal observed in
earlier experiments.
In ref. (\onlinecite{Bacher}) the  PL spectrum
of single semimagnetic nanocrystals show a zero field $\sigma^{+}$ line  with a
full width at half maximum (FWHM) of $\simeq 5$ meV   that red shifts and
narrows upon application  of a magnetic field in the Faraday configuration
\cite{Maksimov,Bacher,Doro}. 
This has motivated  statistical interpretation of the PL spectrum of single
semimagnetic nanocrystals \cite{Bacher,Doro}, linking the
PL linewidth to the  magnetic statistical fluctuations. In such an approach the
PL  FWHM has been shown to scale with $\sqrt{\frac{k_B T}{V}}$. 
On the other
side,   it is apparent that the PL spectrum of a single Mn impurity
\cite{Besombes}  lies in a window of order  
$\Delta_I=\frac{5}{2}\left(|j_h(I)|+|j_e(I)|\right)$, which scales like
$V^{-1}$.  The PL linewidth of fig. 2 also scales with  $\sum_I \Delta_I\propto
V^{-1}$.


 We have explored  whether the quantum approach with 
$N_{Mn}=$3 and 4 can model PL spectra like those of references 
(\onlinecite{Maksimov,Bacher,Doro}).
The justification lies in the distribution of couplings $j_h(I)$, shown 
in fig. 3a for  Cd$_{0.99}$Mn$_{0.01}$Te (dot $\#1$) 
 obtained by random generation of 100  
realizations of the Mn disorder. Such a dot has $N_{Mn}=$28 
and $\langle j_h\rangle=
0.3$ meV. However,  65 $\%$ 
of the Mn  have  $j_h<\langle j_h\rangle$ and, in average, the  4 Mn 
with largest coupling have $j_h>3 \langle j_h\rangle$.
We calculate the
PL taking the 4 Mn atoms with the largest overlap to the carrier wave
function, for a randomly selected Mn disorder realization.
 Their $j_h$ are 1.69, 1.66, 1.1, and 0.45 meV.
In figure 3b we show the corresponding 
$\sigma^{+}$ PL spectrum for values of the magnetic field going from zero to 12 T in the
Faraday configuration. 
We take $\gamma=0.4 meV$ and $T=3 meV$.
As in the experiments, the $\sigma^{+}$ PL red shifts as $B_z$ goes up. 
 The featureless
zero field PL spectrum (with a FWHM of 5 meV) develops,  as $B_z$ increases
narrow structures, as reported experimentally \cite{Doro}.  The narrowing is
also seen in fig. 3c, that  shows  the PL of  dot $\#1$ with 3 Mn atoms, with
values of $j_h$ closer to the average (0.39,0.40 and 0.44 meV), with $T=2$ meV
and $\gamma=0.4$ meV. In fig. 3d we show how the corresponding FWHM  decreases
both upon cooling  and upon increasing $B_z$. The former  is due to the
reduction of the spectral range of the occupied states in the XSM,  resulting
as well  in a red shift thas has been observed experimentally \cite{Bacher}. 
The magnetic field produces a similar effect.

\begin{figure}
[b]
\centering
\includegraphics[width=2.8in]{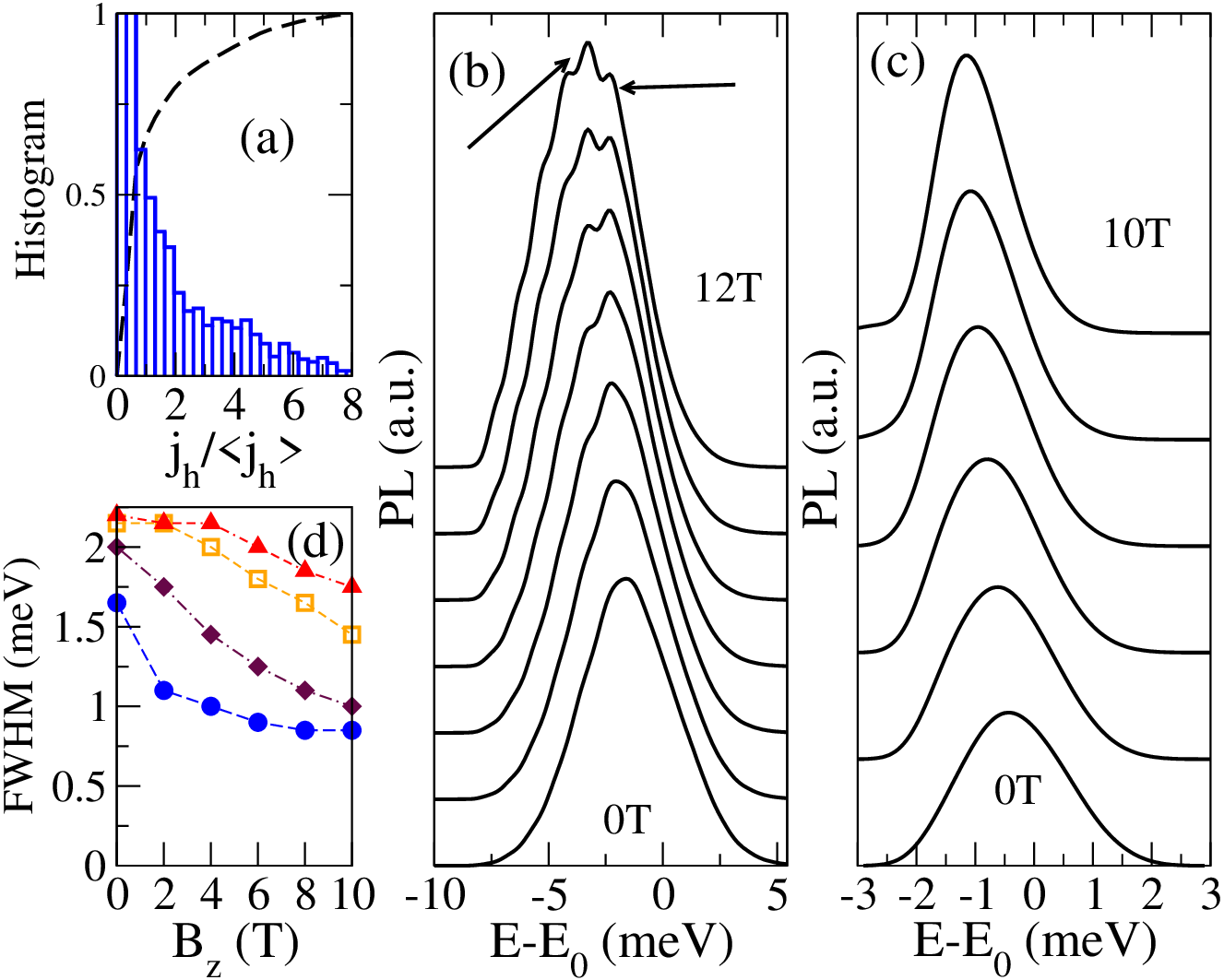}
\caption{ \label{fig3}(Color online).
(a) Normalized histogram of Mn-hole coupling. The dashed line is the integrated
histogram. (b) PL spectrum for 4 Mn impurities (see text) for several
values of $B_z$. (c) Idem for 3 Mn impurities . (d)
FWHM for as a function of $B_z$  and $T$=3,2,1 and 0.5 meV 
(top to bottom).}
\end{figure}

In conclusion, the problem of a few Mn atoms interacting with a single exciton
in a quantum dot is solved via numerical diagonalization of the 
Hamiltonian including electron-Mn, hole-Mn, electron-hole exchange as well as
spin-orbit interactions and the correponding PL spectra are calculated. 
The model accounts for a series of recent reports with the PL of CdTe dots doped
with a single Mn atom and permits to  link the quantum state of one 
Mn spin  to the detection of a photon with a given energy and polarization.
The crucial ingredient for this one on one correspondence is the
anisotropy of the hole-Mn interaction, due to strong spin-orbit interaction for
the holes. The model  reproduce qualitatively the PL spectra of
single dots with presumably tens of  Mn atoms \cite{Maksimov,Bacher,Doro} 
with just 3 and 4 Mn spins strongly coupled to the exciton and with different
coupling strengths. This might indicate that the
width of the  PL spectra coming from a single semimagnetic dot might arise from
a few Mn atoms strongly coupled to the exciton rather than from statistical 
fluctuations of many Mn atoms. 


L. Besombes and H. Mariette are acknowledged for sharing their data before
publication. 
Fruitful discussions with J. J. Palacios and  funding from  
Grant FIS200402356 and the Ramon y Cajal Program (MEC, Spain) are
acknowledged. 
This work  has been partly funded by FEDER funds.



\newpage






\end{document}